\documentclass[graybox, nosecnum]{svmult}

\usepackage{mathptmx}       
\usepackage{helvet}         
\usepackage{courier}        
\usepackage{type1cm}        
\usepackage{makeidx}         
\usepackage{graphicx}        
\usepackage{multicol}        
\usepackage[bottom]{footmisc}
\usepackage{hyperref}        
\usepackage{soul}            
\usepackage{natbib} 
\setcitestyle{round,aysep={}}

\newcommand{\be}{\begin{equation}}
\newcommand{\ee}{\end{equation}}
\newcommand{\bea}{\begin{equation}\begin{aligned}}
\newcommand{\eea}{\end{aligned}\end{equation}}

\def\iso#1#2{\mbox{${}^{#2}{\rm #1}$}}
\def\be1#1{\iso{Be}{1#1}}
\def\n1#1{\iso{N}{1#1}}
\def\c1#1{\iso{C}{1#1}}
\def\ne2#1{\iso{Ne}{2#1}}
\def\na2#1{\iso{Na}{2#1}}
\def\al2#1{\iso{Al}{2#1}}
\def\ar3#1{\iso{Ar}{3#1}}
\def\ca4#1{\iso{Ca}{4#1}}
\def\k4#1{\iso{K}{4#1}}
\def\mn5#1{\iso{Mn}{5#1}}
\def\ni5#1{\iso{Ni}{5#1}}
\def\ge6#1{\iso{Ge}{6#1}}
\def\fe6#1{\iso{Fe}{6#1}}
\def\rb8#1{\iso{Rb}{8#1}}
\def\nb9#1{\iso{Nb}{9#1}}
\def\zr9#1{\iso{Zr}{9#1}}
\def\ru9#1{\iso{Ru}{9#1}}
\def\mo9#1{\iso{Mo}{9#1}}
\def\tc9#1{\iso{Tc}{9#1}}
\def\pd10#1{\iso{Pd}{10#1}}
\def\i12#1{\iso{I}{12#1}}
\def\cs13#1{\iso{Cs}{13#1}}
\def\sm14#1{\iso{Sm}{14#1}}
\def\gd15#1{\iso{Gd}{15#1}}
\def\dy15#1{\iso{Dy}{15#1}}
\def\hf18#1{\iso{Hf}{18#1}}
\def\gd15#1{\iso{Gd}{15#1}}
\def\pb20#1{\iso{Pb}{20#1}}
\def\bi20#1{\iso{Bi}{20#1}}
\def\u23#1{\iso{U}{23#1}}
\def\np23#1{\iso{Np}{23#1}}
\def\pu24#1{\iso{Pu}{24#1}}
\def\cm24#1{\iso{Cm}{24#1}}
\def\th23#1{\iso{Th}{23#1}}
\def\re18#1{\iso{Re}{18#1}}

\def\he#1{\iso{He}{#1}}
\def\li#1{\iso{Li}{#1}}
\def\f1#1{\iso{F}{1#1}}
\def\b1#1{\iso{B}{1#1}}
\def\ba13#1{\iso{Ba}{13#1}}
\def\la13#1{\iso{La}{13#1}}
\def\ta18#1{\iso{Ta}{18#1}}

\def\avg#1{\langle #1 \rangle}

\makeindex             

\begin{document}
\title*{Neutrinos and Heavy element nucleosynthesis}
\author{Xilu Wang \thanks{corresponding author} and Rebecca Surman}
\institute{Xilu Wang \at Institute of High Energy Physics, Chinese Academy of Sciences, Beijing 100049, China, \email{wangxl@ihep.ac.cn}
\and Rebecca Surman \at Department of Physics, University of Notre Dame, Notre Dame, IN 46556, USA, \email{rsurman@nd.edu}}

\maketitle
\abstract{
This chapter discusses three nucleosynthesis processes involved in producing heavy nuclei beyond the iron group that are influenced or shaped by neutrino interactions: the $\nu$ process, the $\nu p$ process and the $r$ process. These processes are all related to explosive events involving compact objects, such as core-collapse supernovae and binary neutron star mergers, where an abundant amount of neutrinos are emitted. The interactions of the neutrinos with nucleons and nuclei through both charged-current and neutral-current reactions play a crucial role in the nucleosynthesis processes. During the propagation of neutrinos inside the nucleosynthesis sites, neutrinos may undergo flavor oscillations that can also potentially affect the nucleosynthesis yields. Here we provide a general overview of the possible effects of neutrinos and neutrino flavor conversions on these three heavy-element nucleosynthesis processes.
}

\section{\textit{Heavy element nucleosynthesis}}

The seminal work of \citet{Burbidge1957} identified three main nucleosynthesis processes responsible for the synthesis of elements heavier than iron: 
the slow ($s$) and rapid ($r$) neutron capture processes and the proton-rich ($p$) process. Since this time, our understanding has evolved and new nucleosynthesis processes have been discovered, \citep[e.g.,][]{Seeger1965,  Arnould1976, Kappeler1999, Arnould2003, Frohlich2006, Travaglio2018, Nishimura2018, solar, Liccardo2018}, yet the basic ideas remain robust. 
In the $s$ process, nuclear species along the center of the valley of stability are built up from iron up to \bi209 by a slow sequence of neutron captures and $\beta$ decays over thousands of years. This process occurs in massive and AGB stars and contributes to about half of the isotopes heavier than iron found in the solar system \citep{Busso1999, Bisterzo2011, Karakas2012}. 
The other 50$\%$ of the solar abundances above iron come from the $r$ process, which synthesizes species on the neutron-rich side of the valley of stability as well as all of the actinides in our universe \citep{Burris2000}. The rapid neutron captures of the $r$ process proceed in timescales of seconds, and various explosive astrophysical events have been suggested for the $r$-process site, such as core collapse supernovae (CCSN), compact binary mergers (neutron star-neutron star or neutron star-black hole mergers, abbreviated NSM in this chapter), collapsars, or more exotic possibilities; see reviews~\citealt{Cowan1991, solar, cowan2021, Kajino2019} and references therein. One common element for most of these possible sites is the presence of an abundant number of neutrinos, suggesting neutrino interactions are of crucial importance to $r$-process nucleosynthesis. 
The small remainder ($\sim0.1-1\%$) of the heavy elements are found on the proton-rich side of stability and most of them can be produced through the charged-particle and photon-induced reactions in explosive nucleosynthesis in supernovae \citep{Woosley1978, Meyer1994, Frohlich2006, solar, Rauscher2010, Kusakabe2011}. A handful of these species possibly owe their origins to neutrino interactions in such environments, through the $\nu$ process \citep{Woosley1990, Heger2005, Suzuki2013, Sieverding2019} and $\nu p$ process \citep{Frohlich2006, Pruet2006, Wanajo2006}. 

Of the aforementioned processes, the $\nu$ process, the $\nu p$ process, and the $r$ process are most strongly shaped by neutrino interactions. The neutrino interactions play a vital role in these nucleosynthesis processes though either neutrino-induced spallation reactions with nuclei ($\nu$ process) or capture reactions on free nucleons:  
\begin{eqnarray}
\nu_e + n \rightleftharpoons p + e^{-},\\
\bar{\nu_e} + p \rightleftharpoons n + e^{+}.
\end{eqnarray}
These capture reactions shape the neutron-to-proton ratio ($n_{n}/n_{p}$) and the electron fraction $Y_{e}=1/(1+n_{n}/n_{p})$ that determine what type of nucleosynthesis is possible in the ejecta, and influence the availability of free nucleons to capture as element synthesis proceeds.

\section{\textit{Neutrino flavor transformations in CCSNe and NSMs}}

A significant fraction of the gravitational binding energy liberated in a CCSN or NSM is carried away by neutrinos, produced thermally and via Eqns.~1 and 2 in the proto-neutron star (PNS) formed in a core-collapse supernova~\citep{Mirizzi+2016} or in the hyper-massive neutron star remnant and accretion disk in a binary merger event~\citep{Cusinato+2022}.
The emitted neutrinos can undergo flavor transformations/oscillations as they propagate outward~\citep{Duan2009, Ruffert1997, Foucart2015}. These neutrino oscillations may change the neutrino spectra to potentially impact the interactions of neutrinos with nucleons and nuclei during the ongoing nucleosynthesis, thus affecting the final nucleosynthesis yields. So it is necessary to know when and where neutrino flavor transformations occur within the nucleosynthesis sites to correctly model heavy-element nucleosynthesis.

For core-collapse supernova explosions, neutrino self interactions dominate the flavor evolution due to the large neutrino fluxes in the vicinity of the PNS, and the resulting collective neutrino oscillations may occur within the radius of $\sim$100 km above the PNS \citep{Duan2010,Chakraborty+2016}. 
In recent years, another neutrino-neutrino interaction through the fast pairwise conversion has been recognized (see reviews~\citealt{Tamborra2020, Capozzi2022} and references therein). These fast flavor conversions are expected to develop on timescales of nanoseconds, which are orders of magnitude smaller than the ``slow" collective neutrino oscillations, and occur at high densities in the proximity of the neutrino decoupling region or in the PNS with length scales of centimeters \citep{Sawyer2005, Sawyer2009, Sawyer2016}. 
When neutrinos travel to the outer layers of the supernovae where the electron number density dominates over neutrino number density, the neutrino-matter resonance Mikheyev-Smirnov-Wolfenstein (MSW) flavor transformations occur \citep{Mikheyev1985, Wolfenstein1978}, corresponding to the baryon density $\rho\sim 10^{2}-10^{3}$ g/cm$^{3}$ \citep{Olive2014}.
In addition to the active neutrino flavors, the existence of sterile neutrinos that do not take part in the weak interactions has been postulated~\citep{Dasgupta+2021}. If sterile neutrinos couple to active flavors, an active-sterile neutrino flavor conversion might happen very close to the neutrinosphere and could greatly reduce the $Y_e$. This could possibly affect the explosion and subsequent nucleosynthesis in supernovae~\citep{Nunokawa1997, McLaughlin1999,Warren+2014,Pllumbi+2015}. 

In a neutron star merger event, an accretion disk forms surrounding the central remnant (a massive neutron star or a black hole) and a neutrino-driven wind is emitted from the hot inner disk after the first few hundred milliseconds~\citep{Surman+2004,Perego2014,Just2015}. Similar neutrino-neutrino interactions (collective neutrino oscillations) should occur in NSMs as its neutrino density is comparable to that of core-collapse supernovae. Additionally, a novel type of neutrino oscillation is possible in this environment due to the protonization of the merger remnant, which leads to an excess of $\bar{\nu_e}$ over $\nu_e$. An almost complete cancellation between the (negative) neutrino-neutrino interaction potential and the matter potential is possible in the merger accretion disk, leading to a matter-neutrino resonance (NMR)~\citep{Malkus2012, Malkus2014, Wu2016, Tian2017, Vlasenko2018, Shalgar2018}. The disk protonization would also possibly result in the fast pairwise conversion above the merger remnant \citep{Wu2017, Wu2017b, Grohs+2022}.

\section{\textit{$\nu$-process nucleosynthesis and neutrinos}}

The $\nu$ process \citep{Woosley1990} occurs in the outer shells of core-collapse supernovae, through the interactions between the neutrinos of all flavors produced from the proto-neutron star and the nuclei present in these shells. The neutrino interactions include charged- (CC) and neutral-current (NC) reactions. Although the neutrino interactions with nuclei are usually negligible, the $\nu$ process can contribute entirely or to a significant portion of the production of a handful of rare isotopes including \li7, \b11, \n15, \f19, \nb92, \tc98, \la138 and \ta180, as well as two radioisotopes \na22 and \al26  \citep{Woosley1990, Heger2005, Byelikov2007, Yoshida2004, Yoshida2005, Yoshida2006, Yoshida2006b, Byelikov2007, Yoshida2008, Hayakawa2008, Hayakawa2010, Nakamura2010, Austin2011, Cheoun2012, Suzuki2013, Rauscher2013, Hayakawa2013, Kajino2014, Lahkar2017, Sieverding2018, Hayakawa2018, Olive2019, Malatji2019, Kusakabe2019, Langanke2019, Li2022, Yao2022}.  

The heavy $\nu$-process isotopes are mainly created though CC reactions. For example, stellar evolution studies \cite{Heger2005} found that the rare odd-odd nuclides \la138 and \ta180 are mainly created by the CC reactions \ba138($\nu_e$, $e^{-}$)\la138 and \hf180($\nu_e$, $e^{-}$)\ta180, thus \la138 and \ta180 could serve as a constraint for $\nu_e$ spectra and luminosities \citep{Hayakawa2008,Byelikov2007,Rauscher2013,Kajino2014,Kheswa2015,Lahkar2017}.
Additionally, the main neutrino reactions to produce \nb92 are the charged-current \zr92($\nu_{e}$, $e^{-}$)\nb92 and the neutral-current \nb93($\nu$($\nu'$),$\nu'$($\nu$)n)\nb92 reactions, and similarly the main reactions producing \tc98 are \mo98($\nu_{e}$, $e^{-}$)\tc98 and \ru99($\nu$($\nu'$),$\nu'$($\nu$)p)\tc98 \citep{Hayakawa2013,Hayakawa2018,Cheoun2012}. 
The light isotopes are also produced through both CC and NC channels. The NC spallation reactions on all species are dominantly induced by $\nu_\mu$ and $\nu_\tau$ neutrinos and their antiparticles (together referred to as $\nu_x$). Thus the $\nu$ process is sensitive to the energies, spectra and luminosity of $\nu_e$ and $\nu_x$ \citep{Woosley1990, Martinez2017, Langanke2019}.

It is expected that the average energies for supernova neutrinos obey an energy hierarchy: $\avg{E_{\nu_e}}$ $<$ $\avg{E_{\nu_x}}$. Based on this assumption, neutrino oscillations should shuffle the lower-energy $\nu_e$ ($\bar{\nu_e}$) spectrum with the higher-energy spectra of $\mu$ or $\tau$ neutrinos (antineutrinos), changing the CC cross section induced by supernova neutrinos. Therefore the $\nu$ process may be a unique probe of the neutrino mass hierarchy or the $\theta_{13}$ mixing angle \citep{Yoshida2006,Yoshida2006b, Grant, Kajino2014, Ko2020}. For example, the light $\nu$ process elements are produced in the outer He layer where there are MSW flavor oscillations corresponding to $\theta_{13}$ mixing~\citep{Yoshida2004, Yoshida2005, Yoshida2006, Yoshida2008}, while the heavy nuclei are not affected by the MSW mechanism as they are synthesized in the O/Ne shell prior to the MSW resonant region \citep{Heger2005}. The observed light isotope ratio \li7$/$\b11 can therefore be used to probe the $\theta_{13}$ mixing angle \citep{Grant}. The more complex effects of neutrino self-interactions/collective neutrino oscillations near the neutrino sphere can have an impact on neutrino nucleosynthesis as well, affecting both the light and heavy $\nu$ process isotopes abundances \citep{Wu2015, Ko2020, Ko2022}. The abundance ratio of heavy to light nuclei, \la138/\b11 is also sensitive to the neutrino mass hierarchy, where solar meteoritic abundances suggest that the normal mass hierarchy is preferred \citep{Ko2020, Ko2022}. 

\section{\textit{$\nu$p-process nucleosynthesis and neutrinos}}

The $\nu p$ process is a recently-identified nucleosynthesis process \citep{Frohlich2006b, Pruet2006, Wanajo2006b}, discovered in the analysis of hydrodynamical simulations of CCSN neutrino-driven winds where most or all of the ejecta was found to be proton rich with low to moderate entropies \citep{Arcones2007,Arnould2003, Hudepohl2010, Janka2016, Fischer2020a}. In general, the $\nu p$ process can occur in explosive environments where proton-rich matter is ejected under the influence of strong neutrino fluxes, producing neutron-deficient isotopes \citep{Frohlich2006b} including the light $p$-nuclei $^{92,94}$Mo and $^{96,98}$Ru \citep{Martinez2014, Pllumbi2015, Wanajo2018, Eichler2018} which 
cannot be produced by the $s$ and $r$ processes nor the classical $p$ process that occurs in the outer part of core collapse supernovae \citep{Dillmann2008, Nishimura2018}.
The $\nu p$-process sites include the inner ejecta of core-collapse supernova (neutrino driven wind; \citealt{Buras2006, Thompson2005, Liebendorfer2008}) and possibly outflows from black hole accretion disks in the collapsar model of gamma-ray bursts \citep{Kizivat+2010}. 

In these proton-rich neutrino-driven winds, matter is initially hot with temperatures well above 10GK and fully dissociated into protons and neutrons. When the matter expands and cools, alpha-rich freeze-out from full nuclear statistical equilibrium (NSE) to quasi-equilibrium (QSE) takes place and nuclei up to \ni56 and even up to \ge64 are produced once the temperature drops below 3 GK. For temperatures between 3 GK and 1.5 GK, sequences of proton-capture reactions and $\beta^{\pm}$ decays dominate.
Without neutrinos, bottleneck (mainly even-even $N$=$Z$) nuclei with long $\beta$-decay half-lives and small proton-capture cross sections impede the synthesis of nuclei beyond the iron peak. In the $\nu p$ process, antineutrino captures on protons produce free neutrons that are easily captured by bottleneck nuclei such as \ge64. The resulting ($n$,$p$) reactions bypass the slow $\beta$-decays, allowing further proton captures to synthesize increasingly heavy nuclei beyond iron \citep{Frohlich2006b, Pruet2006, Thielemann2010}. The $\nu p$-process could therefore be a major contributor to light $p$ nuclei, though its role in galactic chemical evolution remains unclear~\cite{Kobayashi2020}. The $\nu p$-process yields depend on the hydrodynamic state of neutrino-driven winds, nuclear reaction rates \citep{Wanajo2011, Arcones2012, Fujibayashi2015, Bliss2018b, Nishimura2019, Rauscher2019, Jin2020, Vincenzo2021, Sasaki2022}, and the neutrino physics of the astrophysical site.

Collective neutrino oscillations could potentially affect the $\nu p$ process that occurs in CCSN proton-rich winds \citep{Martinez2017, Balantekin2018}. Both the single-angle collective neutrino oscillation calculations of \citealt{Martinez2011} and the multi-angle simulations of \citealt{Sasaki2017} show that collective oscillations act to increase the $\bar{\nu_{e}}$ flux and create a more robust $\nu p$ process. Fast flavor conversions could potentially significantly increase mass loss rates and lead to more proton-rich conditions in neutrino-driven winds, enhancing the possible $\nu p$ process that results in the synthesis of light $p$ nuclei \citep{Xiong2020}. More speculatively, active-sterile neutrino flavor conversion could also help the $\nu p$ process reach heavy elements between Zr and Cd~\citep{Wu2014}.

\section{\textit{$r$-process nucleosynthesis and neutrinos}}

During $r$ process, rapid neutron capture pushes material far from stability and shapes the characteristic abundance pattern with three distinct peaks (at mass numbers $A\sim80$,  $A\sim130$, and $A\sim196$), associated with closed shell structures (at neutron numbers $N=50$, $N=82$, and $N=126$). These features are clearly seen in the abundance pattern of our solar system \citep{Lodders2003, Sneden2008}. The $r$ process is believed to synthesize more than 90$\%$ of heavy elements such as Eu, Au, and Pt found in the solar system \citep{Burris2000}, and it is the only nucleosynthesis path that is responsible for the existence of nuclei heavier than Bi \citep{Liccardo2018}.

The $r$ process occurs when neutron capture rates are much higher than $\beta$-decay rates; this requires an explosive and neutron-rich environment with high densities. However, the most important astrophysical site(s) for the \textit{r}-process are still a subject of debate \citep{cowan2021}. Neutron star mergers have recently been confirmed to be a $r$-process site with the multi-messenger observations of the gravitational event GW170817 \citep{GW170817, Abbott2017, Cowperthwaite2017, Kasen2017}. There are at least two distinct environments for the $r$ process in NSMs:  dynamical ejecta, which is expected to be very neutron-rich (electron fraction $Y_e\sim0.03-0.2$) and robustly produce $r$-process elements all the way to the region of fissioning actinides \citep{Bauswein2013, Hotokezaka2013, Rosswog2013, Endrizzi2016, Lehner2016, Sekiguchi2016, Rosswog2017}; and a viscous and/or neutrino-driven wind from the merger accretion disk, which is believed to be less neutron-rich ($Y_e\sim0.25-0.5$) with the presence of neutrinos \citep{Chen2007, Surman2008, Dessart2009, Wanajo2014, Perego2014, Martin2015, Just2015, Siegel2018}, leading to a weak $r$-process production of elements generally below the second peak. Neutrinos are copiously produced in the merger event and so are expected to play a non-negligible role in setting the electron fraction in at least the neutrino-driven wind component \citep{Caballero2012, Foucart2015, Malkus2016, Kyutoku2018, Fernandez+2019} and possibly more broadly in the ejecta \citep{Wanajo2014,Goriely+2015,Kullmann+2022}.  
Therefore an accurate description of neutrino luminosities and spectra is required to accurately predict light curves and nucleosynthetic yields. Modern merger simulations have progressed from neutrino leakage schemes \citep{Perego2014, Metzger2014, Radice2018, Ardevol2019, Fahlman+2022, Cusinato+2022} to M1 schemes \citep{Just2015, Foucart2015, Fujibayashi2018} and full Monte Carlo transport \citep{Miller+2019}. 
$Y_e$ can also be impacted by the modifications of neutrino and antineutrino spectra due to neutrino flavor transformations. The matter-neutrino resonance has been explored in single-angle and multi-angle approximations \citep{Malkus2012, Foucart2015, Malkus2016, Zhu2016, Frensel2017, Vlasenko2018} along with its potential influence on $r$-process production in the merger disk. Initial studies of fast flavor oscillations in NSMs suggest possible enhancements of $r$-process yields in merger disk ejecta for most cases \citep{Wu2017, Wu2017b, George2020, Li2021, Just2022}.

NSM may be only partially responsible for the Galactic tally of $r$-process elements~\citep{Kyutoku2016, Cote2019, Kobayashi2020, Yamazaki2021}, and exactly how, where, and how much of their contribution to the $r$-process element production has yet to be definitively worked out \citep{Cowan1991, solar, Kajino2019, cowan2021}.
Thus a number of other astrophysical sites have been proposed. In the early 90s a handful of early supernova models predicted the CCSN neutrino-driven wind would be sufficiently high entropy and neutron rich for a robust $r$ process \citep{Meyer+1992,Woosley+1994,Takahashi+1994}. This remained a favored site for a number of years despite early indications that such conditions were difficult to attain \citep{Fuller+1995,McLaughlin+1996,Qian+1996,Hoffman+1997,Otsuki+2000,Thompson+2001,Terasawa+2002}. Modern simulations of CCSN neutrino-driven winds with state-of-the-art neutrino transport predict very similar spectra of $\nu_e$ and $\bar{\nu_e}$ and outflows of modest entropies, such that the neutron-rich and high-entropy conditions needed for a strong $r$ process are not obtained~\citep{Arcones2007, Arcones2011, Fischer2010,Hudepohl2010, Roberts2010, Arcones2011b, Arcones2013, Bliss2018, Akram2020}. Still, a weak $r$ process (producing nuclei up to $A\sim125$) might be possible in this environment, with the ultimate extent of the nucleosynthesis in this site sensitively dependent on neutrino physics~\citep{Fuller+1995,Balantekin2005,Johns2020,Xiong2020}. The effects of neutrino oscillations between active flavors on neutron-rich neutrino-driven wind nucleosynthesis have been explored in, e.g.~\citet{Duan2011,Wu2015, Pllumbi2015, Just2022b}, as well as the active-sterile conversions in~\citet{McLaughlin1999,Beun+2006,Wu2014, Pllumbi2015}.

A rare type of core-collapse event with fast rotation and strong magnetic fields, i.e., magneto-rotational supernovae (MHD SN), may generate neutron-rich outflows though a jet \citep{Nishimura2006, Shibagaki2016, Nishimura2017}. As the neutron-rich material is transported away quickly in this event, the problems associated with long exposures to neutrinos (increasing $Y_{e}$) in the neutrino-driven model can be avoided, and a robust $r$ process can result, though high magnetic fields are required for the nucleosynthesis to extend to the third $r$-process peak~\citep{Winteler2012, mhd, Reichert2021,Reichert2022}. 
A collapsar is a rare core-collapse event linked to long gamma-ray bursts that results in a black hole surrounded by a rapidly-accreting disk of debris. Collapsars were proposed as possible $r$-process sites first in late 90s, when early models of collapsar accretion disks were found to be similar to the neutron-rich accretion disks formed in NSMs \citep{MacFadyen1999}. Semi-analytic work including the effects of neutrino absorption later found that the collapsar disk wind is not sufficiently neutron-rich to allow for the $r$ process \citep{Popham1999, Di2002, Surman2004, McLaughlin2005, Surman2006}. The idea was recently revived when fresh magnetohydrodynamic simulations of the collapsar accretion disk with neutrino radiation in a leakage scheme \citep{Siegel2019} indicated that collapsars can produce sufficiently heavy $r$-process elements to explain observed galactic abundances. This claim has been challenged by simulations with M1 neutrino transport \citep{Just2022} and full frequency-dependent general relativistic neutrino radiation magnetohydrodynamics \citep{Miller2020}, where the predicted nucleosynthesis does not reach the third $r$-process peak. Neutrino oscillations may affect the collapsar disk nucleosynthesis \citep[e.g.,][]{Malkus2012}, similar to their influences on NSM disk ejecta, a topic which is awaiting future exploration.

Other possible $r$-process sites include core-collapse supernovae driven by the quark-hadron transition ~\citep{Fischer2020}, or the primordial black hole-neutron star destruction scenario \citep{Fuller2017}. Additionally, \citealt{Banerjee2011, Banerjee2016} have suggested that MSW flavor transformation in the inverted hierarchy will enhance the amount of neutrons released by \he4 that possibly drive an $r$ process in the He shell of core-collapse supernovae with low metallicities.

\section{\textit{Summary}}

Exactly where and how the heavy elements are synthesized in the Universe remains a mystery despite decades of progress, and neutrinos are an indispensable piece to solve this puzzle. In this chapter, we have reviewed the status of three heavy-element nucleosynthesis processes that are most strongly shaped by neutrino interactions: the $\nu$ process, the $\nu p$ process and the $r$ process. All three are likely to be related to massive stars and explosive events with copious neutrino emissions. We have also summarized the investigations of the neutrino flavor transformations that influence these processes including both active-active neutrino oscillations and active-sterile flavor conversions. With recent discoveries and ongoing explorations of intriguing oscillation phenomena such as collective neutrino oscillations, the matter-neutrino resonance, and fast flavor conversions, the intertwining topics of neutrino physics and nucleosynthesis will continue to be fruitful and active areas of research crucial for the understanding of the astrophysical origins of the heavy elements.

\bibliographystyle{plainnat}

\bibliography{neutrino}

\end{document}